\documentstyle[12pt]{article}
\begin{document}
\topmargin -0.2cm \oddsidemargin -0.2cm \evensidemargin -1cm
\textheight 22cm \textwidth 12cm

\title{ Electromagnetic Light in Medium of Polarized Atoms $^3$He.}  
\author {Minasyan V.N.}

\date{\today}

\maketitle

\begin{abstract}  
First, it is predicted that polarized atoms $^3$He increase a 
value of speed electromagnetic waves. This reasoning implies that
the velocity of electromagnetic waves into gas consisting of 
polarized atoms $^3$He is rather than one in vacuum. 
\end{abstract} 


\vspace{100mm} 

\vspace{5mm}

{\bf 1. INTRODUCTION.} 

\vspace{5mm}

The motivation for theoretical study of the Fermi gas of atoms $^3$He is an attempt at a microscopic understanding of quantization of electromagnetic field. In this letter, we investigate an interaction between electromagnetic field represented as nonideal Bose-gas of elementary Bose-particles of electromagnetic field [1] with Fermi gas of polarized atoms $^3$He. In this context, we show that value of speed   electromagnetic waves is increased by action polarized atoms $^3$He. This effect is appeared by bound electrons into atom.

Theoretical description of quantization local electromagnetic 
field in vacuum within a model of Bose-gas of local electromagnetic waves, which are propagated by speed  $c$ in vacuum, was first proposed by Dirac [2].

\vspace{5mm}

{\bf 11. ANALYSIS}.

\vspace{5mm}  
The starting point our discussion is a model of gas polarized atoms $^3$He representing as the medium consisting $\frac{N}{2}$ neutral polarized atoms $^3$He. Every atom of $^3$He  contains two electrons, therefore, we consider the terms of the interaction between $N$ identical charged electrons with mass $m_e$ and charge $e$ of polarized atoms $^3$He, in a box of volume $V$,  with light particles of electromagnetic field. For beginning, we present the Hamiltonian radiation of electromagnetic light [1], which consists of neutral light bosons with spin one and mass $m$:
$$
\hat{H}_R=\hat{H}_e+\hat{H}_h
$$
\begin{equation} 
\hat{H}_e=
\sum_{ k\leq k_0}\biggl(\frac{\hbar^2
k^2}{2m }+
\frac{mc^2}{2}\biggl )
\vec{E}^{+}_{\vec{k}}\vec{E}_{\vec{k}}- 
\sum_{ k\leq k_0}\biggl (\frac{\hbar^2
k^2}{4m }-\frac{mc^2}{4}\biggl )
\biggl (\vec{E}^{+}_{\vec{k}}
\vec{E}^{+}_{-\vec{k}}+   
\vec{E}_{-\vec{k}}\vec{E}_{\vec{k}}\biggl)
\end{equation}
and

\begin{equation} 
\hat{H}_h=
\sum_{ k\leq k_0}\biggl(\frac{\hbar^2
k^2}{2m }+
\frac{mc^2}{2}\biggl )
\vec{H}^{+}_{\vec{k}}\vec{H}_{\vec{k}}- 
\sum_{ k\leq k_0}\biggl (\frac{\hbar^2
k^2}{4m }-\frac{mc^2}{4}\biggl )
\biggl (\vec{H}^{+}_{\vec{k}}
\vec{H}^{+}_{-\vec{k}}+   
\vec{H}_{-\vec{k}}\vec{H}_{\vec{k}}\biggl)
\end{equation}

where  $\vec { E} ^{+}_{\vec{k}}$, $\vec { H } ^{+}_{\vec{k}}$  and  $\vec {E} _{\vec{k}}$, $\vec {H} _{\vec{k}}$ are, respectively, the second quantzation vectors of wave functions, which are represented as the vector Bose-operators "creation" and "annihilation" of the Bose-particles of electric and magnetic fields with spin one occupying the wave vector $\vec{k}$; $k_0=\frac{mc}{\hbar}$ is the boundary maximal wave number; $c$ is the speed of light. In this context, $\vec {E}^{+}_{\vec{k}}\vec {E}_{\vec{k}}$ and $\vec {H}^{+}_{\vec{k}}\vec {H}_{\vec{k}}$ are, respectively, the scalar operators of the number the Bose-particles of electric and magnetic fields with spin one occupying the wave vector $\vec{k}$.

We now consider the electron gas as an ideal Fermi gas consisting of  $2n$ free charged electrons with mass $m_e$ which is described by the operator Hamiltonian:

\begin{equation}
\hat{H}_0=\sum_{\vec{k},\sigma }\varepsilon_{\vec{k}}
\hat{a}^{+}_{\vec{k},\sigma }\hat{a}_{\vec{k},\sigma} 
\end{equation}
where  $\hat{a}^{+}_{\vec{k},\sigma}$ and 
$\hat{a}_{\vec{k},\sigma }$ are, respectively,  the Fermi operators of creation and 
annihilation for free charged electron with wave-vector $\vec{k}$ and energy $ \varepsilon_{\vec{k}}=\frac{\hbar^2 k^2}{2m_e}$, by the value of its 
spin z-component $\sigma=^{+}_{-}\frac{1}{2}$ which satisfy to the Fermi commutation relations $[\cdot\cdot\cdot]_{+}$ as:

\begin{equation}
\biggl[\hat{a}_{\vec{k},\sigma}, \hat{k}^{+}_{\vec{p}^{'},
\sigma^{'}}\biggl]_{+} =
\delta_{\vec{k},\vec{k^{'}}}\cdot\delta_{\sigma,\sigma^{'}}
\end{equation}

\begin{equation}
[\hat{a}_{\vec{k},\sigma}, \hat{a}_{\vec{k^{'}}, \sigma^{'}}]_{+}= 0
\end{equation}

\begin{equation}
[\hat{a}^{+}_{\vec{k},\sigma}, \hat{k}^{+}_{\vec{p^{'}}, 
\sigma^{'}}]_{+}= 0
\end{equation}

We now are aimed to describe the property of the model a light boson gas-charged electron gas mixture confined in a box of volume $V$.  In this context, the main part of the  Hamiltonian of a light boson gas-charged electron gas mixture consists of the term of the Hamiltonian of the light Bose-particles $\hat{H}_R$ and the term of the Hamiltonian of an ideal Fermi charged electron gas $\hat{H}_0$ as  well as the term $\hat{H}_Q $ of the interaction between the density of the light boson modes  and the density of the charged electron modes: 

\begin{equation}
\hat{H}=\hat{H}_R+\hat{H}_e +\hat{H}_Q
\end{equation}

To present the Hamiltonian  $\hat{H}_Q$ of the interaction 
between  light boson modes and charged electron modes, we introduce the method of second quantization for system of $N$ fermions. Thus, we may rewrite down the second quantization wave functions for one charged electron in point of coordinate  $\vec{r}$ in following form:

\begin{equation}
\psi (\vec{r}, \sigma)= 
\frac{1}{\sqrt{V}}\sum_{\vec{k}}\hat{a}_{\vec{k},\sigma } 
e^{i \vec{k} \vec{r}}
\end{equation}

\begin{equation}
\psi^{+} (\vec{r}, \sigma)= 
\frac{1}{\sqrt{V}}\sum_{\vec{k}}\hat{a}^{+}_{\vec{k},\sigma } 
e^{-i \vec{k} \vec{r}}
\end{equation}

and

\begin{equation}
\int \psi^{+} (\vec{r}, \sigma) \psi (\vec{r}, \sigma) dV= \sum_{k,\sigma}\hat{a}^{+}_{\vec{k},\sigma}
\hat{a}_{\vec{k},\sigma}=\hat{N}
\end{equation}

In beginning, we consider the term  $H_Q$ between light boson modes and an charged electron modes:

\begin{equation} 
\hat{H}_Q =\int \sum_{\sigma}\psi^{+} (\vec{r},
\sigma) H_q \psi (\vec{r}, \sigma)dV 
\end{equation}

where $\vec {p}=-i\hbar\nabla$; $m_e$ and $ e $ are, respectively, a mass and a charge of charged electron; $A$ is the vector potential of electromagnetic light. However, due to introduction quantized vector electric and magnetic fields in [1]

\begin{equation} 
\vec {E}=- \frac{\hbar \sqrt{2\pi}}{c\sqrt{m}}\cdot
\frac{d {\vec{H}_0}}{d t}+ c\sqrt{2m\pi}\cdot \vec {E}_0
\end{equation} 
and
\begin{equation} 
\vec {H}=\frac{\hbar \sqrt{2\pi}}{c\sqrt{m}}\cdot
\frac{d {\vec{E}_0}}{d t} + c\sqrt{2m\pi} \vec{H}_0
\end{equation}

we can rewrite the Hamiltonian of interaction between one 
electron with electromagnetic field as

\begin{eqnarray} 
H_q &=&\frac{1}{m_e}\biggl[\biggl(\vec {p} - 
\frac{e\hbar \sqrt{2\pi}\vec{E}_0}{c\sqrt{m}}
\biggl)^2-\vec {p}^2\biggl]+ \nonumber\\
&+&\frac{1}{ m_e}\biggl[\biggl(\vec {p} + 
\frac{e\hbar \sqrt{2\pi}\vec{H}_0}{c\sqrt{m}}
\biggl)^2-\vec {p}^2\biggl] 
\end{eqnarray}

where $\vec {E}_0$ and $\vec {H}_0$ are presented via the Bose operators $\vec {E}^{+}_{\vec{k}}$, $\vec {H}^{+}_{\vec{k}}$ and $\vec {E}_{\vec{k}}$, $\vec {H}_{\vec{k}}$:

\begin{equation} 
\vec {E}_0= \frac{1}{V}\sum_{\vec{k}}\biggl(
\vec {E} _{\vec{k}} e^{i(
\vec{k}\vec{r} + kc t )} +\vec {E}^{+}_{\vec{k}}
e^{-i(\vec{k}\vec{r} + kc t)}\biggl)
\end{equation} 

and
\begin{equation} 
\vec {H}_0= \frac{1}{V}\sum_{\vec{k}}\biggl(
\vec {H} _{\vec{k}} e^{i(
\vec{k}\vec{r} + kc t )} +\vec {H}^{+}_{\vec{k}}
e^{-i(\vec{k}\vec{r} + kc t)}\biggl)
\end{equation}

Inserting  (16) and (17) in (15) by taking into consideration (10)-(12), and also
 
\begin{equation}
\frac{1}{V}\int e^{i\vec{k}\vec{r}} dV=\delta_{\vec{k}}
\end{equation}

we may rewrite down:
$$
\hat{H}_Q=\hat{H}_{Q,e}+\hat{H}_{Q,h}
$$

\begin{eqnarray}
\hat{H}_{Q,h}&=&\frac{\pi \hbar^2e^2N}{m m_e c^2 V}
\sum_{k_1,\sigma}\sum_{k_2,\sigma}
\sum_{\vec{k_3 }}\sum_{\vec{k_4 }}\hat{a}^{+}_{\vec{k}_1,\sigma}
\hat{a}_{\vec{k}_2,\sigma}\biggl(\vec{H} _{\vec{k}_3}  +
\vec {H}^{+}_{-\vec{k}_3}\biggl) \times 
\nonumber\\
&\times &\biggl(\vec{H}   _{\vec{k}_4}  +
\vec {H}^{+}_{-\vec{k}_4}\biggl)
\delta_{\vec{k}_2+\vec{k}_3+\vec{k}_4-\vec{k}_1}-
\nonumber\\
&- &\frac{e\hbar \sqrt{2\pi} }{m_e \sqrt{m} c}\sum_{k_1,\sigma}\sum_{k_2,\sigma}
\sum_{\vec{k_3 }}\vec{k}_2\hat{a}^{+}_{\vec{k}_1,\sigma}
\hat{a}_{\vec{k}_2,\sigma}\biggl(\vec{H} _{\vec{k}_3}  +
\vec {H}^{+}_{-\vec{k}_3}\biggl) \delta_{\vec{k}_2+
\vec{k}_3-\vec{k}_1}
\end{eqnarray}
which has an analogy form for operator $\hat{H}_{Q,e}$.

Hence, we note that the Fermi operators $\hat{a}_{\vec{k},\sigma}$,  $\hat{a}_{\vec{k},\sigma }$ communicates with the Bose operators $\vec {E}^{+}_{\vec{k}}$,
$\vec {E}_{\vec{k}}$ and $\vec {H}^{+}_{\vec{k}}$,
$\vec {H}_{\vec{k}}$ because they are  in depended.

The introduction of the random - phase approximation [3], proposed by Bohm-Pines,  has a following form:

$$
\sum_{\vec{k}_1}\hat{a}^{+}_{\vec{k}_1,\sigma}\hat{a}_{\vec{k}_1-\vec{k},\sigma}\approx\sum_{\vec{k}_1}\hat{a}^{+}_{\vec{k}_1,\sigma}\hat{a}_{\vec{k}_1-\vec{k},\sigma}\delta_{\vec{k},0}=\sum_{\vec{k}}
\hat{a}^{+}_{\vec{k},\sigma}\hat{a}_{\vec{k},\sigma}=N
$$
In this respect, an application of later in (19), we obtain the term of operator $\hat{H}_Q $:

\begin{eqnarray} 
\hat{H}_Q &=&\frac{\pi \hbar^2e^2 N}{m m_e c^2 V}\sum_{\vec{k}}
\biggl(\hat{E}_{\vec{k}}+\hat{E}^{+}_{-\vec{k}}\biggl)
\biggl(\hat{E} _{-\vec{k}} +\hat{E}^{+}_{\vec{k}}\biggl)+
\nonumber\\
&+&\frac{\pi \hbar^2e^2 N}{m m_e c^2 V}\sum_{\vec{k}}
\biggl(\hat{H}_{\vec{k}}+\hat{H}^{+}_{-\vec{k}}\biggl)
\biggl(\hat{H} _{-\vec{k}} +\hat{H}^{+}_{\vec{k}}\biggl)
\end{eqnarray}

Thus, the main part of the  Hamiltonian of a light boson gas, by application $\hat{H}_R$ by (1), (2), reduces to a following form:
$$
\hat{H}=\hat{H}_a+\hat{H}_b
$$

where

\begin{eqnarray} 
\hat{H}_a &=&\sum_{\vec{k}}\biggl (\frac{\hbar^2k^2}{2m}+\frac{mc^2}{2}+\frac{2\pi \hbar^2e^2N}{m m_e c^2 V}\biggl )
\hat{E}^{+}_{\vec{k}}\hat{E}_{\vec{k}}-  
\nonumber\\
&-&\sum_{\vec{k}}\biggl (\frac{\hbar^2k^2}{2m}-\frac{mc^2}{2}-\frac{\pi \hbar^2e^2N}{m m_e c^2 V}\biggl )
\biggl (\hat{E}^{+}_{\vec{k}}
\hat{E}^{+}_{-\vec{k}}+   
\hat{E}_{-\vec{k}}\hat{E}_{\vec{k}}\biggl)
\end{eqnarray}

\begin{eqnarray} 
\hat{H}_b &=&\sum_{\vec{k}}\biggl (\frac{\hbar^2k^2}{2m}+\frac{mc^2}{2}+\frac{2\pi \hbar^2e^2N}{m m_e c^2 V}\biggl )
\hat{H}^{+}_{\vec{k}}\hat{H}_{\vec{k}}-  
\nonumber\\
&-&\sum_{\vec{k}}\biggl (\frac{\hbar^2k^2}{2m}-\frac{mc^2}{2}-\frac{\pi \hbar^2e^2N}{m m_e c^2 V}\biggl )
\biggl (\hat{H}^{+}_{\vec{k}}
\hat{H}^{+}_{-\vec{k}}+   
\hat{H}_{-\vec{k}}\hat{H}_{\vec{k}}\biggl)
\end{eqnarray}

The evaluation of energy levels of the operators $\hat{H}_a$ and $\hat{H}_b$ within diagonal form, we apply new linear transformation of a vector Bose-operator which is similar to the Bogoliubov transformation for a scalar Bose-operator [4]:

\begin{equation}
\vec {E}_{\vec{k}}=\vec {H}_{\vec{k}}=\frac{\vec {h}_{\vec{k}} + 
M_{\vec{k}}\vec {h}^{+}_{-\vec{k}}} {\sqrt{1-M^2_{\vec{k}}}}
\end{equation} 

where $M_{\vec{k}}$ is the real symmetrical functions  
from  a wave vector $\vec{k}$.

Thus, the operator Hamiltonian $\hat{H}$ takes a following form:  
\begin{equation}
\hat{H}=
2\sum_{ k\leq k_0}\eta_{\vec{k}}\vec {h}^{+}_{\vec{k}} 
\vec {h}_{\vec{k}}+ \sum_{\vec{k},\sigma }\varepsilon_{\vec{k}}
\hat{a}^{+}_{\vec{k},\sigma }\hat{a}_{\vec{k},\sigma}
\end{equation}
\begin{eqnarray}
\eta_{\vec{k}}&=&\sqrt{\biggl (\frac{\hbar^2k^2}{2m}+\frac{ mc^2}{2} +\frac{2\pi \hbar^2e^2N}{m m_e c^2 V}\biggl )^2-\biggl (\frac{\hbar^2k^2}{2m}-\frac{ mc^2}{2}-\frac{2\pi \hbar^2e^2N}{m m_e c^2 V}\biggl )^2}=
\nonumber\\
&=&\hbar k v
\end{eqnarray}
where $v$ is the velocity of photon in charged electron medium, which equals to

\begin{equation}
v=c\sqrt{1+\frac{4\pi \hbar^2e^2N}{m^2 m_e c^4 V}}
\end{equation}
In the letter [1], it is found the mass of the boson of an electromagnetic field
which represents as a new fundamental constant:

\begin{equation}
m=\frac{m_e e^4}{2\hbar^2 c^2}=2.4 \cdot 10^{-35} kg
\end{equation}   

An application of later leads to 

\begin{equation}
v=c\sqrt{1+\frac{16\pi a^3 N}{V}}
\end{equation}
where $a=\frac{\hbar^2 }{m_e e^2}$ is the Bohr radius. 

Introducing the parameter 
$$
r_s =\biggl(\frac{3V}{4\pi N}\biggl)^{\frac{1}{3}}
\frac{ m_e e^2}{\hbar^2}
$$ 
which characters a density of atomic gas, and then, we obtain 

\begin{equation}
v=c\sqrt{1+\frac{4}{r^3_s}}>c
\end{equation}

Thus, we proved that velocity of electromagnetic waves into gas consisting polarized atoms $^3$He is rather than one in vacuum, due to an interaction between electrons of atom and electromagnetic field. This fact may have significant output for application in instruments and devices, which work on the photo effect.

\newpage 
\begin{center} 
{\bf References} 
\end{center} 
 
\begin{enumerate} 
%\bibitem{1}
\item 
V.N.~Minasyan~.," Superfluid Component of Electromagnetic Field  and New Fundamental Constant in The Nature", arXiv:0903.0223, ~(2009)~,  [physics.gen-ph], 
V.N.~Minasyan~.," Light Bosons of Electromagnetic Field  and  Breakdown of Relativistic Theory" ~(2009)~;  
arXiv:0808.0567v10 [physics.gen-ph]
%\bibitem{2} 
\item 
P.A.M..~Dirac~, "The Principles of Quantum Mechanics", ~Oxford at the  
Clarendon  press  
(1958), "Lectures on Quantum Mechanics". ~Yeshiva University New York~  
(1964)  
%\bibitem{3}
\item
N.N.~Bogoliubov~, Jour. of Phys.(USSR), ~{\bf 11},~23~(1947)
%\bibitem{4}
\item
D.~Bohm~, D.~Pines~., Phys.Rev.B.~{\bf 82},~625~(1951); D.~Bohm~,
D.~Pines~, Phys.Rev.B.~{\bf 92},~609~(1953).

\end{enumerate} 
\end{document}